\documentclass[twocolumn]{aastex62}

\received{January 1, 2018}
\revised{January 7, 2018}
\accepted{\today}
\submitjournal{ApJ}

\shorttitle{Jets in PNe}
\shortauthors{Guerrero, Rechy-Garc\'\i a \& Ortiz}

\begin{document}

\title{
Space Velocity and Time Span of Jets in Planetary Nebulae}

\correspondingauthor{Mart\'\i n A. Guerrero}
\email{mar@iaa.es}

\author[0000-0002-7759-106X]{Mart\'\i n A. Guerrero}
\affil{Instituto de Astrof\'isica de Andaluc\'ia, IAA-CSIC \\
Glorieta de la Astronom\'\i a s/n, 18008 Granada, Spain}

\author{Jackeline Suzett Rechy-Garc\'\i a}
\affiliation{Instituto de Astrof\'isica de Andaluc\'ia, IAA-CSIC \\
Glorieta de la Astronom\'\i a s/n, 18008 Granada, Spain}
\affiliation{Instituto de Astronom\'\i a, Universidad Nacional Aut\'onoma de 
M\'exico \\
Apdo.\ Postal 70264, 04510, Ciudad de M\'exico, Mexico}

\author{Roberto Ortiz}
\affiliation{Escola de Artes, Ciências e Humanidades, USP \\ 
Av.\ Arlindo Bettio 1000, 03828-000 São Paulo, Brazil}



\begin{abstract}

Fast highly-collimated outflows including bipolar knots, jet-like features, 
and point-symmetric filaments or string of knots are common in planetary 
nebulae (PNe).  
These features, generally named as jets, are thought to play an 
active role in the nebular shaping immediately before or at the
same time that fast stellar winds and D-type ionization fronts
shock and sweep up the nebular envelope.
The space velocity, radial distance from the central star and kinematic 
age of jets in PNe cannot be determined because the inclination angle
with the line-of-sight is usually unknown.  
Here we have used the large number of jets already detected
in PNe to derive orientation-independent properties from a
statistical point of view.  
We find that jets in PNe can be assigned to two different populations:
a significant fraction ($\simeq$70\%) have space velocities below 100
km~s$^{-1}$, whereas only $\simeq$30\% have larger velocities.
Many jets move at velocities similar to that of their
parent PNe and are found close to the nebular edge.
We propose that these jets have been slowed down in their interaction
with the nebular envelope, contributing to the expansion of their PN.
The time span before a jet dissolves is found to be generally
shorter than 2,500 yrs.
Since most jets are found in young PNe of similar (1,000--3,000 yrs)
age, it can be concluded that jets are mostly coeval with their PNe.

\end{abstract}



\section{Introduction} \label{sec:intro}

A fraction of planetary nebulae (PNe) presents morphological and
kinematic features indicative of fast and highly collimated outflows.
These are mostly detected in narrow-band images \citep[e.g.,][]{C96b} 
as low-ionization features with a variety of morphologies, including
jet-like features, bipolar compact knots and point-symmetric filamentary
structures or strings of knots.  
Such morphological variety has made researchers to coin 
a number of names to designate them, including BRETs
\citep[Bipolar Rotating Episodic jeTs,][]{LVR1995},
FLIERs \citep[Fast Low-Ionization Emission Regions,][]{BRT_etal1993},
or simply jets.
We will adopt hereafter the latter term to refer
to them, despite their morphological diversity.  
Kinematically, they are characterized by anomalous radial velocities
with respect to the velocity field of the PN and narrow, unresolved
velocity structures.

Jets are found in PNe spanning almost all morphological
classes \citep{GVL99,GCM01} and evolutionary stages
\citep{A00,CLHetal2000,BL90,GGCG2004}.  
It is nowadays commonly accepted that jets play a critical role 
in the shaping of axisymmetric PNe \citep{ST98}, although the 
first detection of such a fast collimated outflow in a PN, namely 
NGC\,2392 \citep{GBS85}, was completely unexpected because it was 
unforeseen that progenitor stars of PNe could host accretion 
disks and/or have strong magnetic fields to provide the required 
conditions for the collimation and acceleration of a jet \citep{L99}.
Different models have been proposed to explain the formation of 
jets in PNe, including hydrodynamical focusing \citep{FBL96},
magnetic collimation \citep{GS97}, and accretion disks
\citep{M87,SL94}, but none of them offer a comprehensive 
explanation to all their properties \citep{B98,GCM01}. 

Most observational studies of jets report the spatial and kinematic 
properties in one single object or a small sample of sources
\citep[e.g.,][among many others]{BPI87,MS92,LMP1993,GVL99,C00,AL2012}.
Only a few studies have dealt with averaged properties for
a sample of PNe, considering the morphological and kinematic
properties \citep{GCM01,AG2016} or the linear momenta of the  
jets \citep{TMW2014}.   
The real space velocity of a jet could provide basic 
insights on its formation mechanisms and interactions
with the nebular envelope, but it is unknown due to
projection effects because the angle between the
velocity vector and the line of sight cannot be determined.  
Only in a few cases there is some additional information that can be 
used to constrain the real space velocity.  
The width of the line at zero intensity, $FWZI$, is a direct measurent 
of the real space velocity \citep{HRH87}. 
Unfortunately, this procedure requires a high signal-to-noise
ratio in order to determine the zero intensity level accurately.  
More frequently, the real space velocity of the collimated
outflow is worked out after a spatio-kinematic model has
been built for the main nebula \citep[e.g.,][]{CGLetal2010}
or for multiple ejecta located at different orientations
\citep[e.g.,][]{MGT99}, or the shock-velocity is derived
from suitable spectral diagnostics \citep[e.g.,][]{GMC2004}.
This requires adopting simple assumptions on the geometry and kinematics
of the jets and their relation with different nebular structures.

In this paper we present an analysis of the spatial and 
kinematic properties of a sample of jets in PNe using a
statistical approach.
The observed distributions of radial velocity and distance to 
the central star (CSPN) projected on the plane of the sky have
been modeled to derive the intrinsic distributions of space
velocity and distance to the CSPN.
This has allowed us to infer sound conclusions about their velocity 
and age distributions and compare them to those of their parent PNe.

We describe the selection of the sample and discuss
possible selection effects and biases in $\S$2.  
The distributions of the observed radial velocities and projected
distances to the CSPNe of the jets are modeled in $\S$3 to infer
their actual space velocity and distance distributions, and to
investigate their ages.  
Jets in PNe are found to belong to two different populations
with different kinematic properties.
In the framework of these results, we discuss
different formation mechanisms of jets in PNe
and the effects of their interactions with the
nebular envelope in $\S$4. 
The main results and conclusions are summarized in $\S$5.

\section{The Sample of Jets} 
\label{sec:sample}

We have compiled in Table~\ref{tab:charac1} a list of 58 PNe with 
jets (columns 1 and 2) for which there are available high-dispersion
spectroscopic optical observations that provide their kinematic
information.  
Although additional jets are suggested in optical and 
near-infrared narrow-band images of a larger number
of PNe \citep[e.g.,][]{SMV2011}, we have disregarded
those with no available kinematic information.  
Similarly, proto-PNe with jets
\citep[e.g., CRL\,618 or Hen\,3-1475, ][]{Lee13,Cox03,BH01}
have not been included in this study because some of their
nebular properties such as expansion velocity and linear
distance to the CSPN cannot be derived properly.
Finally, there is a notorious sample of bipolar hourglass-shaped PNe
that are not included in Table~\ref{tab:charac1} neither for they do
not comply with the morphological and kinematic definition of a jet,
even if they present highly collimated lobes
\citep[e.g., M\,2-9, ][]{Clyne_etal15}.

The final sample of \emph{bonafide} jets in these 58
sources amounts to 85, as some PNe present several
independent sets of such outflows.  
We investigate next the kinematic and spatial properties of these
PNe and their jets.

\subsection{Properties of the PNe with Jets} 
\label{subsec:sample1}

The basic properties (distance, angular size, and expansion velocity)
of the PNe with jets are compiled in Table~\ref{tab:charac1}.
The distances ($d$) to these sources listed in column 3 were
adopted from \citet{FPB16} when available, otherwise adopted 
from \citet{CKS92}.  
The angular sizes of their semi-major ($a$) and semi-minor ($b$)
axes are listed in columns 4 and 5, respectively.  
These were measured mostly in H$\alpha$ $\lambda$6563 \AA\ 
\emph{Hubble Space Telescope} (\emph{HST}) WFPC2/PC images.
If no available, other \emph{HST} filters or instruments, our own H$\alpha$ 
images \citep[for instance, The IAC Morphological Catalog of Northern 
Galactic PNe,][]{M96}, or publicly available H$\alpha$ images \citep[for 
instance, The University of Hong Kong/Australian Astronomical 
Observatory/Strasbourg Observatory H-alpha Planetary Nebula database, HASH 
PN,][]{PBF2016} were used for this purpose as indicated in the footnotes 
of Table~\ref{tab:charac1}.  
Sizes were estimated using a contour at a 10\% level of the peak
of the nebular emission.
Finally, the expansion velocities of the main nebular shells 
($v_{\rm PN}$) as given in the literature are listed in column 6.

\startlongtable
\begin{deluxetable*}{llcccclcccl}
\tablenum{1}
\tabletypesize{\scriptsize}  
\tablecaption{Properties of PNe and their jets}
\tablecolumns{11}
\tablewidth{0pt}
\tablehead{
\multicolumn{6}{c}{\underline{~~~~~~~~~~~~~~~~~~~~~~Planetary Nebulae and Their Properties~~~~~~~~~~~~~~}} & 
\multicolumn{4}{c}{\underline{~~~~~~~~~Jets and Their Properties~~~~~~~~~~}} & 
\colhead{} \\
\multicolumn{1}{l}{Name}  &
\multicolumn{1}{l}{PN\,G} & 
\multicolumn{1}{c}{$d$}   & 
\multicolumn{1}{c}{$a$}   & 
\multicolumn{1}{c}{$b$}   & 
\multicolumn{1}{c}{$v_{\rm PN}$} & 
\multicolumn{1}{l}{ID}    & 
\multicolumn{1}{c}{$v_{\rm r}$} & 
\multicolumn{1}{c}{$\theta_{\rm s}$} & 
\multicolumn{1}{c}{$x$}   & 
\multicolumn{1}{l}{References} \\ 
\multicolumn{2}{c}{} & 
\multicolumn{1}{c}{(kpc)} & 
\multicolumn{2}{c}{(arcsec)} & 
\multicolumn{1}{c}{(km~s$^{-1}$)} &
\multicolumn{1}{c}{} &
\multicolumn{1}{c}{(km~s$^{-1}$)} &  
\multicolumn{1}{c}{(arcsec)} & 
\multicolumn{2}{c}{} 
}
\startdata
IC\,4634\tablenotemark{a}          & 000.3$+$12.2 &   2.8  &   4.5 &   2.9 &   18   & A-A'           &  19   &  10.2 &  2.05 & 1,2 \\
                                   &              &        &       &       &        & D-D'           &  19   &   4.5 &  0.90 & \\
IC\,4776\tablenotemark{b}          & 002.0$-$13.4 &   4.4  &   5.0 &   2.4 & $<$8   & A-B            &  49.6 &   4.7 &  1.34 & 3,4 \\ 
                                   &              &        &       &       &        & C-D            &  84.1 &   8.1 &  2.31 & \\
M\,1-37                            & 002.6$-$03.4 &  14.4  &   1.4 &   0.9 &   11.1 & PA=129$^\circ$ &   7.6 &   2.0 &  1.16 & 3 \\
Hb\,4\tablenotemark{a}             & 003.1$+$02.9 &   2.9  &   4.2 &   3.0 &   21.5 & N-S            & 160   &  14.6 &  2.11 & 5 \\ 
M\,3-15\tablenotemark{a}           & 006.8$+$04.1 &   5.5  &   2.1 &   1.6 &   15   & E-W            &  90   &   5.6 &  2.13 & 6,7 \\
M\,2-42\tablenotemark{b}           & 008.2$-$04.8 &   9.6  &  11.2 &   6.4 &   15   & PA=0$^\circ$   &  15   &  10.0 &  2.22 & 7 \\
M\,1-32\tablenotemark{c}           & 011.9$+$04.2 &   3.6  &   5.6 &   4.9 &   15   & $\dots$   & 180   &   0.0 &  0.00 & 6 \\
HuBi\,1                            & 012.2$+$04.9 &   3.5  &   9.3 &   8.3 &   22   & $\dots$   & 165   &   0.0 &  0.00 & 8 \\ 
M\,2-40                            & 024.1$+$03.8 &   5.4  &   2.7 &   2.3 &   17.6 & PA=88$^\circ$  &  16.0 &   4.8 &  1.64 & 3 \\
Pe\,1-17\tablenotemark{b}          & 024.3$-$03.3 &   6.2  &   7.2 &   2.2 &   24   & PA=19$^\circ$  &  14.5 &   6.0 &  1.20 & 9 \\ 
                                   &              &        &       &       &        & PA=50$^\circ$  &   5.5 &   4.8 &  0.95 & \\
IC\,4593\tablenotemark{a}          & 025.3$+$40.8 &   1.6  &   7.8 &   5.6 &   15   & A-B            &   2   &  12.5 &  1.59 & 10 \\
                                   &              &        &       &       &        & C              &   1   &   7.4 &  0.94 &   \\
IC\,4846                           & 027.6$-$09.6 &   7.1  &   1.7 &   1.4 & $\dots$& A1-A2          &  48   &   4.6 &  1.74 & 11 \\
                                   &              &        &       &       &        & B1-B2          &  13   &   3.1 &  1.18 & \\
NGC\,6751\tablenotemark{a}         & 029.2$-$05.9 &   2.5  &  11.8 &  11.5 &   41.8 & Slit f         &  31   &  17.7 &  1.49 & 12 \\
PC\,19\tablenotemark{a}            & 032.1$+$07.0 &   8.3  &   0.7 &   0.7 &   30.5 & PA=28$^\circ$  &  35   &   2.3 &  0.95 & 9 \\
M\,1-66                            & 032.7$-$02.0 &   6.4  &   1.6 &   1.4 &   20.0 & PA=131$^\circ$ &   7.0 &   3.5 &  1.94 & 3 \\
NGC\,6741\tablenotemark{a}         & 033.8$-$02.6 &   3.3  &   4.1 &   2.6 &   23.4 & A-B            &  22.7 &   8.0 &  1.78 & 3 \\
                                   &              &        &       &       &        & C              &   7.3 &   6.3 &  1.40 & \\
NGC\,6778\tablenotemark{d}         & 034.5$-$06.7 &   2.8  &  10.0 &   4.5 &   26   & PA=47$^\circ$  & 192   &  32.0 &  2.56 & 13 \\
                                   &              &        &       &       &        & PA=15$^\circ$  & 113   &  34.3 &  2.74 & \\
NGC\,6572                          & 034.6$+$11.8 &   1.5  &   4.2 &   2.9 &   14   & PA=15$^\circ$  &   8   &   6.5 &  1.04 & 9,14 \\
                                   &              &        &       &       &        & PA=162$^\circ$ &  39   &   9.0 &  1.43 & \\
NGC\,7009\tablenotemark{a}         & 037.7$-$34.5 &   1.3  &  14.3 &   5.6 &   20.8 & Slit a         &   6.5 &  26.9 &  1.89 & 15 \\ 
NGC\,6210\tablenotemark{e}         & 043.1$+$37.7 &   2.1  &   6.6 &   5.2 &   34.2 & A              &  19.5 &   8.4 &  1.11 & 3 \\
                                   &              &        &       &       &        & B              &  30.6 &   4.7 &  0.62 & \\
                                   &              &        &       &       &        & C              &  29.4 &  17.6 &  2.36 & \\
Hen\,2-429\tablenotemark{b}        & 048.7$+$01.9 &   3.7  &   3.6 &   2.8 &   30.6 & PA=90$^\circ$  &   5.9 &   7.6 &  2.16 & 3 \\
Hu\,2-1\tablenotemark{a}           & 051.4$+$09.6 &   4.2  &   1.5 &   0.8 &   15   & C1-C2          &  54   &   3.6 &  1.44 & 16,17 \\
                                   &              &        &       &       &        & C3-C4          &  58   &   3.0 &  1.20 & 17 \\
                                   &              &        &       &       &        & D              &   2   &   5.1 &  2.00 & 16,17 \\
Abell\,63\tablenotemark{b}         & 053.9$-$03.0 &   2.4  &  50.0 &  22.3 &   17   & polar caps     &   5.5 & 142.4 &  5.70 & 18 \\
NGC\,6891\tablenotemark{a}         & 054.1$-$12.1 &   2.9  &   6.9 &   6.8 &   10   & A-A'           &  14   &   6.0 &  0.76 & 14,19 \\
Necklace\tablenotemark{f}          & 054.2$-$03.4 &   4.6  &   7.3 &   3.3 &   28   & PA=34$^\circ$  &  54   &  57.3 &  6.59 & 20 \\
Hen\,1-1                           & 055.3$+$02.7 &   5.7  &   3.8 &   2.6 &   33   & PA=315$^\circ$ &  43.5 &   6.1 &  1.32 & 9 \\
K\,3-35\tablenotemark{c}           & 056.0$+$02.0 &  10.1  &   2.0 &   0.6 &   10   & PA=7$^\circ$   &  20   &   1.8 &  0.79 & 21,22 \\
M\,2-48\tablenotemark{b}           & 062.4$-$00.2 &   6.4  &  15.6 &   7.4 &   10   & B1-B2          &   7   & 109.1 &  9.48 & 23,24 \\
BD+30$^\circ$3639\tablenotemark{a}  & 064.7$+$05.0 &   2.2  &   3.5 &   3.2 &   35.5 & Slit NS        &  90   &   5.5 &  1.17 & 25,26 \\ 
ETHOS\,1\tablenotemark{b}          & 068.1$+$11.1 &   8.7  &   9.7 &   9.7 &   55   & NW             &  60   &  30.0 &  4.00 & 27 \\
NGC\,6881\tablenotemark{a}         & 074.5$+$02.1 &   3.6  &   2.2 &   1.4 &   14   & PA=127$^\circ$ &   9   &  13.8 &  2.11 & 28 \\
NGC\,6884\tablenotemark{a}         & 082.1$+$07.0 &   3.2  &   3.2 &   2.9 &   19   & A-A'           &  22   &   2.8 &  0.77 & 29 \\
                                   &              &        &       &       &        & NEK-SWK        &  36   &   4.8 &  1.30 & \\
NGC\,6826\tablenotemark{a}         & 083.5$+$12.7 &   1.4  &  13.3 &  10.5 &    6   & FLIER          &  44.5 &  13.9 &  1.10 & 15,30,31 \\
Hu\,1-2\tablenotemark{d}           & 086.5$-$08.8 &   5.1  &   4.8 &   2.1 &   32   & NW-SE          &  60   &  27.6 &  4.06 & 32 \\
NGC\,6543\tablenotemark{e}         & 096.4$+$29.9 &   1.2  &  13.0 &   8.3 &   16   & F-F'           &  32   &  15.3 &  1.35 & 15,33,34 \\
                                   &              &        &       &       &        & J-J'           &  39   &  21.2 &  1.88 & \\
NGC\,7662\tablenotemark{a}         & 106.5$-$17.6 &   1.4  &  14.9 &  13.2 &   25   & Knot \#2       &  30   &  17.2 &  1.25 & 35 \\
                                   &              &        &       &       &        & Jet \#22       &  70   &  19.4 &  1.41 & \\
NGC\,7354\tablenotemark{e}         & 107.8$+$02.3 &   1.3  &  13.9 &  10.9 &   28   & A1-A5          &   5   &  20.3 &  1.24 & 36 \\
KjPn\,8\tablenotemark{c}           & 112.5$-$00.1 & $\dots$&   3.2 &   2.9 &   16   & A1-A2          & 220   & 120~~ &  0.91 & 37 \\
                                  &              &        &       &       &        & C1-C2          &   0   & 420~~ &  3.19 & \\
K\,4-47\tablenotemark{c}           & 149.0$+$04.4 &  10.4  &   4.4 &   1.0 & $\dots$& PA=41$^\circ$  &  59   &   3.8 &  0.90 & 38 \\
IC\,2149                           & 166.1$+$10.4 &   2.8  &   5.4 &   2.2 &   24   & PA=67$^\circ$  &   0   &   5.1 &  0.83 & 39 \\
J\,320                             & 190.3$-$17.7 &   5.8  &   4.5 &   2.8 &   16.0 & PA=348$^\circ$ &  19.9 &   9.8 &  2.39 & 3 \\
                                   &              &        &       &       &        & PA=338$^\circ$ &  34.0 &   7.8 &  1.89 & \\
                                   &              &        &       &       &        & PA=306$^\circ$ &  24.8 &   3.7 &  0.85 & \\
NGC\,2392\tablenotemark{e}         & 197.8$+$17.3 &   1.4  &  20.1 &  20.1 &  120   & PA=70$^\circ$  & 180   &  22   &  0.98 & 40 \\
NGC\,1360\tablenotemark{c}         & 220.3$-$53.9 &   0.6  & 229.2 & 169.2 &   24   & Slit c         &  76   & 318   &  1.27 & 15,41 \\
                                   &              &        &       &       &        & Slit d         &  81   & 420   &  1.68 & \\
M\,1-16\tablenotemark{b}           & 226.7$+$05.6 &   6.2  &  18.2 &  10.2 & $\dots$& Slit b         & 252.5 &  59   &  4.74 & 15,42 \\
M\,3-1\tablenotemark{a}            & 242.6$-$11.6 &   4.5  &   5.0 &   3.2 &   24.5 & A-B            &  14.9 &  12.4 &  2.39 & 3 \\
K\,1-2\tablenotemark{g}            & 253.5$+$10.7 &   3.6  &  65.2 &  54.8 &   25   & A3             &  20.5 &  16.7 &  0.56 & 43 \\
Wray\,17-1\tablenotemark{b}        & 258.0$-$15.7 &   2.3  &  46.8 &  37.8 &   28   & A-B            &  23.5 &  32   &  0.97 & 43 \\
Wray\,17-1\tablenotemark{b}        &              &        &       &       &        & C-D            &  13   &  22.5 &  0.68 & \\
NGC\,3242\tablenotemark{a}         & 261.0$+$32.0 &   1.0  &  12.0 &   7.9 &  25-30 & FLIER          &  25   &  15.6 &  1.26 & 30,31 \\
Hen\,2-47                          & 285.6$-$02.7 &   3.8  &   2.2 &   2.1 &   11.0 & PA=0$^\circ$   &  23.5 &   4.0 &  1.33 & 3 \\
                                   &              &        &       &       &        & PA=64$^\circ$  &  22.9 &   3.9 &  1.28 & \\ 
Fg\,1\tablenotemark{b}             & 290.5$+$07.9 &   1.5  &  32   &  26.3 &   36   & A-A'           &  75   & 110   &  2.75 & 44,45 \\
NGC\,3918\tablenotemark{e}         & 294.6$+$04.7 &   1.6  &   8.2 &   5.8 &   22.6 & A-C            &  27   &  23.2 &  2.72 & 43 \\
                                   &              &        &       &       &        & B              &  10   &  13.5 &  1.59 & \\
Hen\,2-90\tablenotemark{a}         & 305.1$+$01.4 &   4.9  &   1.4 &   1.2 & $\dots$& Knot h         &  26   &  10   &  7.25 & 46 \\
MyCn\,18\tablenotemark{a}          & 307.5$-$04.9 &   2.8  &   1.8 &   1.2 &   24   & Knots 2,18     & 460   &  49.9 &  8.32 & 47 \\
NGC\,5307\tablenotemark{a}         & 312.3$+$10.5 &    3.2 &   9.2 &   5.4 &   24.5 & PA=0$^\circ$   &   6.8 &   6.6 &  1.00 & 48 \\
                                   &              &        &       &       &        & PA=32$^\circ$  &  12.0 &   5.3 &  0.80 & \\
                                   &              &        &       &       &        & PA=60$^\circ$  &  12.1 &   3.6 &  0.55 & \\
                                   &              &        &       &       &        & PA=134$^\circ$ &   7.3 &   5.4 &  0.81 & \\
                                   &              &        &       &       &        & PA=166$^\circ$ &  14.7 &   8.2 &  1.24 & \\
Hen\,2-115                         & 321.3$+$02.8 &    5.0 &   1.4 &   1.1 &   13.4 & A-B            &   5.9 &   2.9 &  2.59 & 3 \\
Hen\,2-186\tablenotemark{a}        & 336.3$-$05.6 &    6.9 &   1.3 &   0.8 & $\dots$& A-B            & 135   &   4.5 &  4.59 & 38 \\
NGC\,6337\tablenotemark{b}         & 349.3$-$01.1 &    1.5 &  28.0 &  29.1 & $\dots$& A-B            &  50   &  45.8 &  1.57 & 38 \\
M\,1-26                            & 358.9$-$00.7 &    2.1 &   1.7 &   1.5 & $<$7   & PA=82$^\circ$  &  38.2 &   3.8 &  2.76 & 3 \\
                                   & 358.9$-$00.7 &        &       &       &        & PA=145$^\circ$ &  47.7 &   3.4 &  2.51 & \\                           
\enddata
\tablenotetext{a}{\emph{HST} WFPC2-PC F658N [N~{\sc ii}] $\lambda$6584 \AA\ image.} 
\tablenotetext{b}{No \emph{HST} image available.}
\tablenotetext{c}{Image adopted from the reference in the last column.}
\tablenotetext{d}{NOT ALFOSC [N~{\sc ii}] $\lambda$6584 \AA\ image.}
\tablenotetext{e}{\emph{HST} WFPC2-WFC F658N [N~{\sc ii}] $\lambda$6584 \AA\ image.}
\tablenotetext{f}{\emph{HST} WFC3/UVIS images.}
\tablenotetext{g}{\emph{HST} WFPC2-WFC F502N [O~{\sc iii}] $\lambda$5007 \AA\ image.}
\tablecomments{
(1) \citet{GMC2004}, 
(2) \citet{GMRetal2008}, 
(3) \citet{RG2019}, 
(4) \citep{Sowicka2017}, 
(5) \citet{LSM97}, 
(6) \citet{RG17}, 
(7) \citet{AL2012}, 
(8) Guerrero et al.\ 2019, in prep., 
(9) \citet{GVL99}, 
(10) \citet{C96}, 
(11) \citet{MGT01}, 
(12) \citet{CGLetal2010}, 
(13) \citet{GM2012}, 
(14) \citet{AG2016}, 
(15) \citet{Letal2012}, 
(16) \citet{M95}, 
(17) \citet{M01}, 
(18) \citet{Mitchell_etal2007}, 
(19) \citet{GMMV00}, 
(20) \citet{CSMetal2011}, 
(21) \citet{M00b}, 
(22) \citet{BGMetal2014}, 
(23) \citet{VLMetal2000}, 
(24) \citet{LLEetal2002}, 
(25) \citet{BM99}, 
(26) \citet{B00}, 
(27) \citet{MCBetal2011}, 
(28) \citet{GM98}, 
(29) \citet{MGT99}, 
(30) \citet{BPI87}, 
(31) \citet{B98}, 
(32) \citet{MBGR2012}, 
(33) \citet{MS92}, 
(34) \citet{R99}, 
(35) \citet{AGR2017}, 
(36) \citet{CVM2010}, 
(37) \citet{LRMetal2002}, 
(38) \citet{C00}, 
(39) \citet{VMTetal2002}, 
(40) \citet{GLSR2012}, 
(41) \citet{GGCG2004}, 
(42) \citet{CS1993}, 
(43) \citet{C99}, 
(44) \citet{LMP1993},
(45) \citet{P96}, 
(46) \citet{Getal01}, 
(47) \citet{O00}, and
(48) Rechy-Garc\'\i a et al., in preparation.  
}
\label{tab:charac1}
\end{deluxetable*}

\begin{figure}[ht!]
\includegraphics[bb=28 68 580 740,width=1.0\columnwidth]{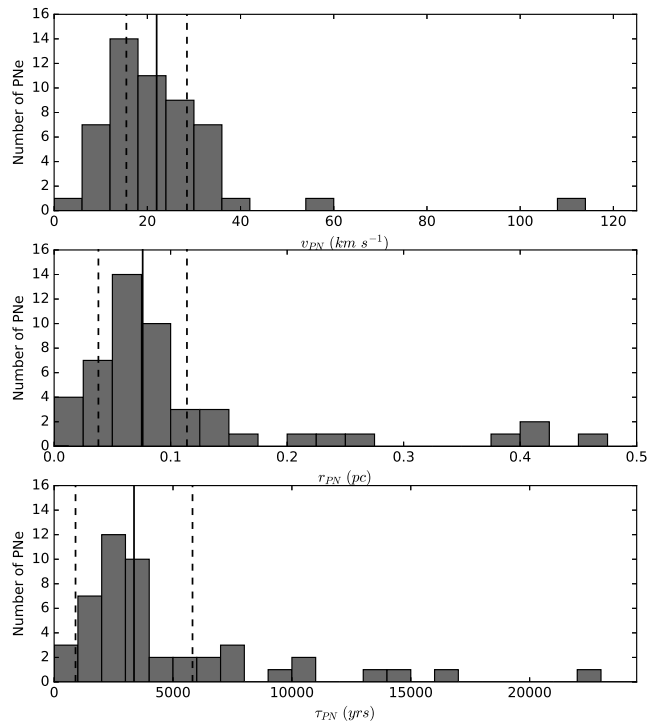}
\caption{
Distributions of the expansion velocity (\emph{top}), linear radius 
(\emph{middle}), and kinematic age (\emph{bottom}) of the sample 
of PNe with jets.  
In each panel, the vertical solid line marks the median value and 
the vertical dashed lines the 1-$\sigma$ standard deviation.  
\label{pn.properties}}
\end{figure}

In Figure~\ref{pn.properties} we present the distributions of
the expansion velocity ($v_{\rm PN}$), linear nebular radius 
($r_{\rm PN}$) 
computed as 
\begin{equation}
  r_{\rm PN} = \frac{\sqrt{a^2+b^2}}{\sqrt{2}} \times d, 
\end{equation}
and kinematic age ($\tau_{\rm PN}$)
defined as 
\begin{equation}
  \tau_{\rm PN} = r_{\rm PN} / v_{\rm PN}  
\end{equation}
of the PNe in Table~\ref{tab:charac1}.  
These distributions indicate that PNe with jets have median values of 22$\pm$7
km~s$^{-1}$ for their expansion velocities and 0.076$\pm$0.038 pc for their
radii.
The distribution of the kinematic ages implies a median value of
3400$\pm$2500 yrs, which points to relatively young PNe.
It must be noted, however, that kinematic ages are affected by many
dynamical effects which make untrustable its use to estimate the PN
age \citep{VMG2002,CSSetal2007}.
Rather, we can use the nebular size of these PNe to estimate 
their ages, as theoretical models of PN formation show that
the nebular size increases with age with varying rates for
different initial mass of its progenitor \citep[e.g.,][]{TA2016}.  
Accordingly, an age between 1000 and 3000 yrs can 
be estimated from the linear size of 0.076 pc.

The relatively small nebular size of PNe with jets suggests
that these are found mostly in relatively young PNe, and may
reveal a bias that hampers the detection of old, evolved
outflows that will be discussed in \S\ref{subsec.large_bias}.

\subsection{Properties of the jets}

\subsubsection{Radial velocities}
\label{subsec:sample2}

Columns 7 and 8 of Table~\ref{tab:charac1} provide the identification
of the 85 jets with kinematic information of these 58 PNe and their
absolute systemic radial velocities,
$\vert v_{\rm r}^{\rm jet}\vert \equiv v_{\rm r}$, respectively.  
The identifications correspond to the names or labels used in the
corresponding original references (column 11) or to their position
angle (PA) on the sky.
The systemic radial velocities, i.e., the velocities along the
line of sight relative to the systemic velocity of the PN, have
been mostly adopted from the literature, as given in the reference
listed in column 11.  
These velocities are typically measured using the [N~{\sc ii}] $\lambda$6584 
\AA\ emission line rather than the H$\alpha$ or [O~{\sc iii}] $\lambda$5007 
\AA\ emission lines, because the smaller thermal width of the [N~{\sc ii}] line
allows a more accurate determination of the velocity, and the intensity of the
[N~{\sc ii}] lines are enhanced in these low-excitation features with respect
to that of H$\alpha$ and [O~{\sc iii}] \citep{B94,GCM01}. 
In many cases (e.g, M\,1-16 and NGC\,1360) we have complemented the
available kinematic information with measurements we have carried
out on echelle spectra downloaded from the San Pedro Martir (SPM)
Kinematic Catalogue of Galactic Planetary Nebulae \citep{Letal2012}.

\begin{figure*}
\includegraphics[width=2.0\columnwidth]{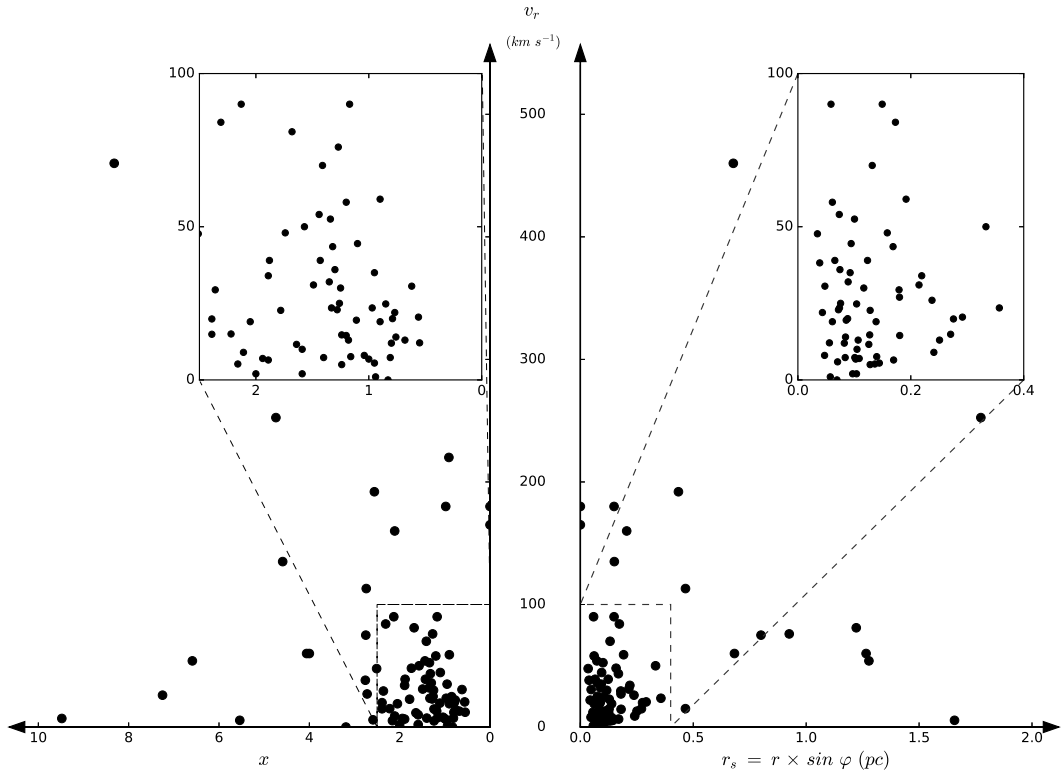}
\caption{
Distribution of the radial velocity of collimated outflows $v_{\rm r}$ versus
the relative distance to the CSPN with respect to the nebular radius $x$ as
defined in Eq.~\ref{eq.x} (\emph{left}) and versus the distance of the collimated outflow to the CSPN
projected onto the plane of the sky $r_{\rm s}$ (\emph{right}).  
The innermost regions of these plots are zoomed on the insets.
\label{pos_vr}}
\end{figure*}

Many jets in PNe exhibit complex kinematic and 
morphological structures consisting of several
components.  
In particular, a steady increase of velocity with distance to the 
center of the PN is reported for a significant fraction of them 
\citep[e.g., Hb\,4,][]{Derlopa_etal2019}.  
In order to use a consistent definition for their radial velocity, 
we adopted $v_{\rm r}$ to be the largest measured velocity of every 
single jet in each PN.  
Different situations can be described.  
In PNe with a single pair of compact knots confined to one orientation 
(e.g., NGC\,7009), we adopted $v_{\rm r}$ as the semi-difference between 
their radial velocities.  
Similar criterium was applied to PNe with two or more pairs of 
independent knots along different orientations (e.g., KjPn\,8
and NGC\,6572) for each pair of jets.
On the other hand, some PNe exhibit jets with 
noticeable changes in orientation and projected
velocity.  
In those cases, wherever the jet was detected as a continuous ejection (e.g.,
PC\,19) or as a set of discrete knots (e.g., Fg\,1), we adopted for $v_{\rm r}$
the maximum semi-difference between the radial velocities of spatially
opposite components.  
Finally, there are cases where only one single component at one side 
of the PN is detected or there is no available kinematic information
for its counterpart (e.g., the feature B in NGC\,3918).  
Here, we determined $v_{\rm r}$ as the difference between the radial
velocity of the jet and the PN systemic radial velocity.

\subsubsection{Distances of the jets to the CSPNe}
\label{subsec:sample3}

Table~\ref{tab:charac1} also compiles the information on the angular
distance $\theta_{\rm s}$ from the tip of the jet to the CSPN (column
9).  
These values have been measured mostly in [N~{\sc ii}] images,
either in available \emph{HST} or archival images, or adopted
from the original references \citep[e.g., Fg\,1,][]{P96}.

There are two PNe in our sample, namely M\,1-32 and HuBi\,1, that can be
classified as ``spectroscopic bipolar nebulae'' \citep[][Guerrero et al.,
in prep.,]{RG17}, i.e., their jets have been discovered spectroscopically,
but are not spatially resolved from the CSPN ($\theta_{\rm s}$=0).
There are also cases such as those of MyCn\,18 \citep{B97}
and Hen\,2-90 \citep{SN00,Getal01} for which the jet consists
of several knots aligned along the same direction.  
Following the prescription given above, $\theta_{\rm s}$ was assumed to be
the angular distance between the CSPN and the farthest knot (the tip) of
the jet.

It can be interesting to refer these angular distances $\theta_{\rm s}$
to the nebular radii, particularly because the detection of a jet may
be compromised when its projection falls onto the nebular shell whose
bright emission may outshine it and hinder its detection.  
The relative distance to the CSPN of the jet with
respect to the nebular radius has been defined as
\begin{equation}
x = \theta_{\rm s} / \theta_{\rm PN-jet},
\label{eq.x}
\end{equation}
where $\theta_{\rm PN-jet}$ is the radius of the nebular shell at 2\% level 
from the peak intensity along the direction of the jet.
These values are compiled in column 10 of Table~\ref{tab:charac1}.

\begin{figure}
\includegraphics[width=1.00\columnwidth]{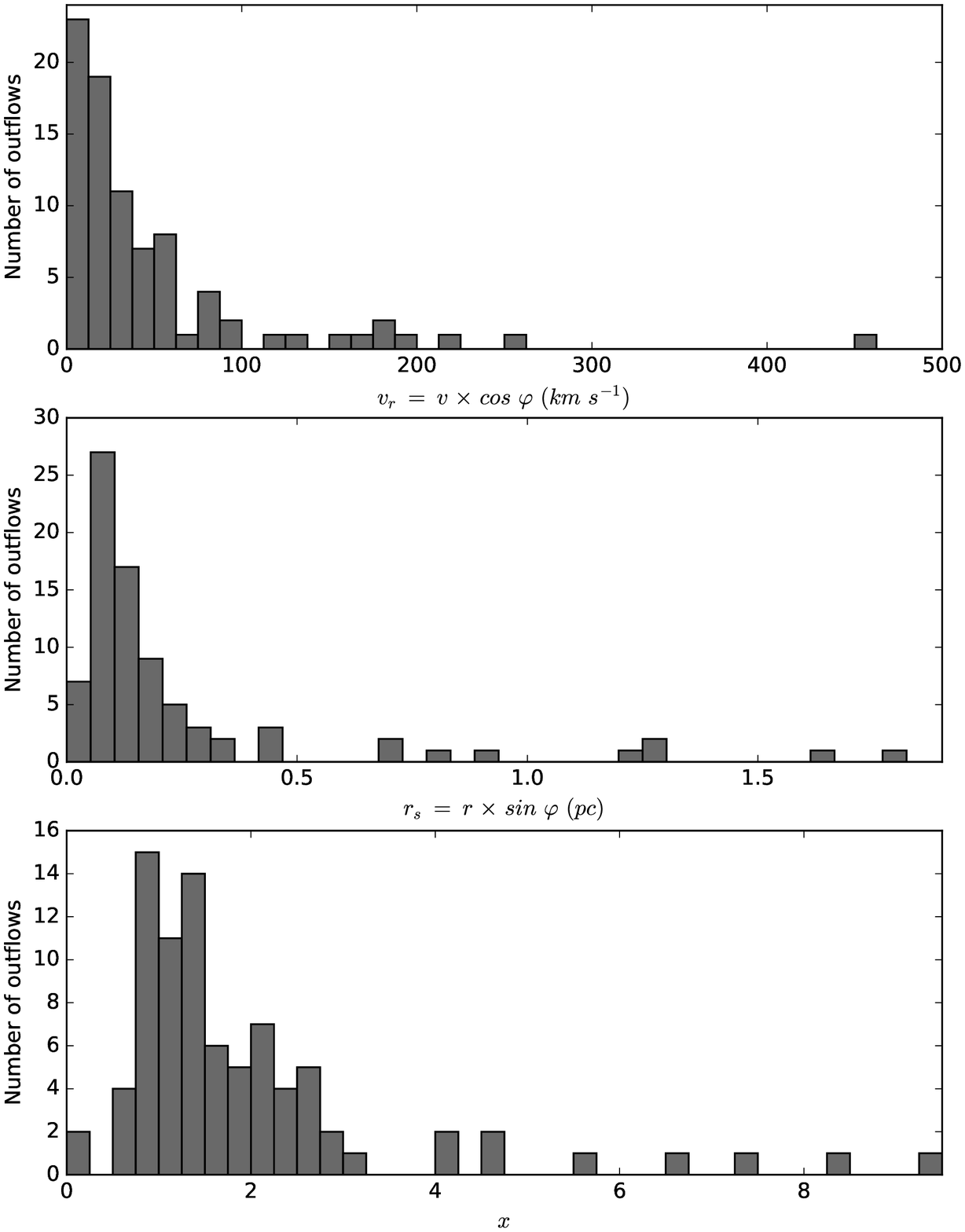}
\caption{
Distributions of the absolute radial velocity $\vert v_{\rm r}\vert$
(\emph{top}), distance to the CSPN projected onto the plane of the
sky $r_{\rm s}$ (\emph{middle}), and relative position with respect
to the nebular radius $x$ (\emph{bottom}) of the jets listed in
Table~\ref{tab:charac1}.
}
\label{histo}
\end{figure}

\subsection{Biases}
\label{sec.bias}

Several biases may hinder the detection of a
jet in a PN and affect our sample selection.  
These are considered below.

\subsubsection{Jets projected onto the PN}
\label{subsec.small_bias}

Jets in PNe at small inclination angles with the line-of-sight
would be projected onto the main nebular shell, making difficult
their detection as they are embedded in bright nebular emission.
Their enhanced brightness in emission lines of low-ionization species and
their high degree of collimation, however, make them show up conspicuously
on narrow-band [N~{\sc ii}] images \citep{M96}, on H$\alpha$+[N~{\sc ii}]
to [O~{\sc iii}] ratio maps \citep{C96b}, or on high-dispersion spectroscopic
observations \citep{MS92,BM99}, even if the surrounding nebular emission is
much brighter.  
Still, there can be a lack of detections of jets projected onto the
nebular shells of compact and presumably young PNe, as testified by
the small number of ``spectroscopic bipolar nebulae'' and young
($\lesssim1,000$ yrs) PNe with jets.
Spectroscopic searches for high velocity outflows
projected close to CSPNe can improve this situation.

\subsubsection{Jets far from the PN} 
\label{subsec.large_bias}

As jets move away from the main nebulae, they are expected
to expand and interact with the circumstellar medium.
As they disperse, the detection of low-surface brightness clumps will become
difficult.
The time scale for jets to disappear will be discussed
in \S\ref{subsec.times}, but it is obvious that old jets
will be found at large projected distances from their
parent PNe, particularly if they have also large expansion
velocities and move close to the plane of the sky.
As a result, old and fast outflows might be 
missed in searches using images with small
field of view.  
Indeed, there is only a small fraction of jets with angular distances to the
CSPN greater than 100$^{\prime\prime}$ (Fg\,1, KjPn\,8, M\,2-48, and NGC\,1360),
projected distances to the CSPN greater than 0.5 pc (Ethos\,1, Fg\,1, Hu\,1-2,
M\,1-16, M\,2-48, MyCn\,18, NGC\,1360, the Necklace Nebula), or relative size
with respect to their parent nebulae along the direction of the outflow greater
than 5 
(Hen\,2-90, M\,2-48, MyCn\,18, and the Necklace Nebula).
Searches for high velocity outflows using large-scale
narrow-band images can improve this situation.

\section{Results}

The data on the radial velocity and distance to the CSPN of the
jets compiled in Table~\ref{tab:charac1} have been plotted in
Figure~\ref{pos_vr}.
The left panel shows the absolute value of their systemic radial velocities
($v_{\rm r}$) versus their relative distances to the CSPN with respect to the
nebular radius ($x$). 
This panel reveals that most jets are located in
the region $x<2.5$ and $v_{\rm r}<100$ km~s$^{-1}$.
This region is zoomed in the inset in this plot to reveal a
notorious lack of data points for relative distances $x<0.5$.

Meanwhile, the right panel of Figure~\ref{pos_vr} compares
$v_{\rm r}$ with the linear distance to the CSPN of the jet
projected onto the plane of the sky ($r_{\rm s}$).
The latter is derived from the distance to the nebula and the angular
distance of the jet to the CSPN $\theta_s$ as
\begin{equation}
  r_{\rm s} \simeq \theta_s \times d.  
\label{eq.rsky}
\end{equation}
This panel indicates that most jets have projected distances to the CSPN
between 0.05 and 0.3 pc, as emphasized in the inset shown in this panel.

In Figure~\ref{pos_vr}, dots close to the horizontal axes somehow
correspond to jets that move close to the plane of the sky, whereas
dots at high $v_{\rm r}$ might be associated with those moving close
to the line of sight.  
Of course, the above statements are to be questioned, as the location 
of data points in these plots depends on the space velocity, linear
distance to the CSPN (i.e., age), and inclination with respect to the
line of sight of the jets.  
The right panel of Figure~\ref{pos_vr} actually shows that jets
with radial velocities $\geq 100$ km~s$^{-1}$ tend to have relatively 
small ($\lesssim$0.5 pc) projected distances to the CSPN, but the one 
of M\,1-16.  
If we assume these small projected distances are due to 
small inclination angles with the line of sight, this 
suggests that jets have expansion velocities generally 
not much higher than 200 km~s$^{-1}$.  
To allow an investigation of these from a statistical point of view,
the occurrence of $v_{\rm r}$, $r_{\rm s}$, and $x$ of the jets compiled
in Table~\ref{tab:charac1} are plotted in Figure~\ref{histo}.  
These will be discussed in the next sections.

\subsection{Distribution of the Absolute Systemic Radial Velocities of Jets}

The distribution of $v_{\rm r}$ in the top panel of Figure~\ref{histo}
shows the number of jets for each radial velocity bin.  
The $v_{\rm r}$ distribution declines to velocities
$\simeq$100 km~s$^{-1}$, with most jets having
radial velocities $\leq$ 100 km~s$^{-1}$.
Faster outflows, $v_{\rm r} \geq$ 100 km~s$^{-1}$, are less
frequent ($\simeq$12\%) and appear spread over a wide range 
of velocities.

\subsubsection{Implications for the Space Velocity Distribution}

Let us define $\varphi$ as the inclination angle with respect to the line of
sight of a jet and $v_{\rm exp}^{\rm jet} \equiv v$ its space (i.e., unprojected)
velocity.  
Then, the radial velocity listed in column 8 of Table~\ref{tab:charac1} 
corresponds to 
\begin{equation}
v_{\rm r} \equiv |v_{\rm r}^{\rm jet}| = 
  v_{\rm exp}^{\rm jet} \times \vert \cos\varphi \vert \equiv v \times
  \vert \cos\varphi \vert.
  \label{eq.vr}
\end{equation}
Let us define the distribution of the space velocities $v$ as Q($v$)
and that of the observed absolute radial velocities $v_{\rm r}$ as
P($v_{\rm r}$).  
It is possible to derive information on Q($v$) from P($v_{\rm r}$)
on the assumption that the spatial orientation of jets in PNe is
isotropic\footnote{
\citet{CAM98} presented convincing arguments that the orientations
of axisymmetric PNe in the Galaxy is isotropic, but we must be 
cautious to extend this to jets in PNe as their formation 
mechanism may differ from the general processes involved in the
formation of axisymmetric PNe.
}.  
Let further assume that the detection of a jet in a PN is not
affected by any observational bias, i.\ e., that the probability
of detecting it {\it does not depend on its spatial orientation
or its projection onto the nebula or its angular distance to the
nebula} (see \S\ref{sec.bias}).  
Then, it follows that the space distribution of jets
$f(\varphi,\psi)$ with the inclination angle $\varphi$
from the line of sight and the azimuthal angle $\psi$
fulfills the relation: 
\begin{equation}
\int_{0}^{+\pi}\int_{0}^{2\pi} f(\varphi,\psi) d\psi \sin \varphi d\varphi = 1. 
\end{equation}
Since the space distribution of jets does not
depend on the spatial direction, then it follows: 
\begin{equation}
f(\varphi,\psi) \int_{0}^{\pi}\int_{0}^{2\pi} d\psi \sin \varphi d\varphi = 4 \pi \; f(\varphi,\psi) 
\end{equation}
for a constant value of the space distribution
\begin{equation}
  f(\varphi,\psi) = \frac{1}{4 \pi}.
  \label{eq_iso}
\end{equation}

\begin{figure*}[ht!]
\begin{center}
\includegraphics[bb=31 318 583 711,width=1.85\columnwidth]{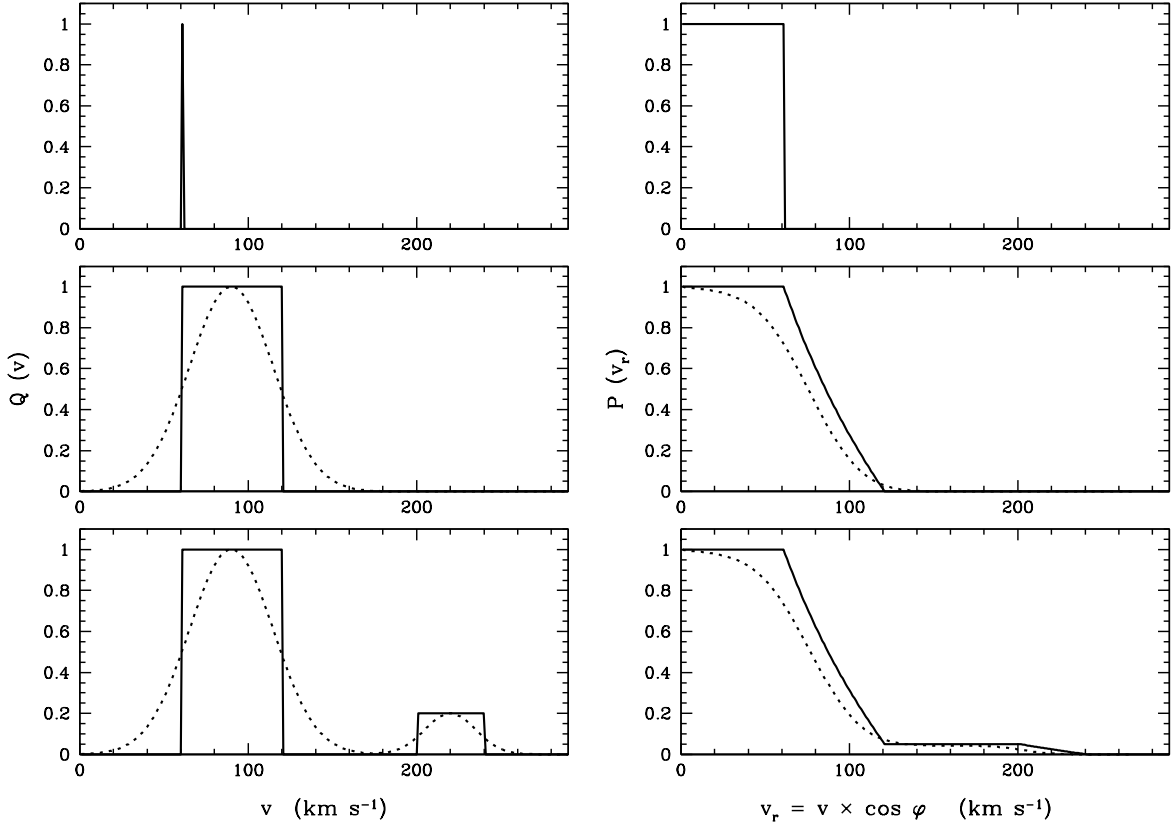}
\end{center}
\caption{
(\emph{left}) Different distributions of space velocities Q($v$) and 
(\emph{right}) their corresponding distribution of radial velocities 
P($v_{\rm r}$).  
The Q($v$) distributions consist of a single space velocity at 
$v$=60 km~s$^{-1}$ in the top panels, a single top-flat (solid 
line) and Gaussian (dashed line) component centered at 
$v$=90 km~s$^{-1}$ in the middle panels, and two top-flat (solid 
line) and Gaussian (dashed line) components centered at $v$=90 
km~s$^{-1}$ and $v$=220 km~s$^{-1}$ in the bottom panels.  
In the latter, the high-velocity component has lower probability than the 
low-velocity component.  
}
\label{vr_theory}
\end{figure*}

\begin{figure*}[ht!]
\begin{center}
\includegraphics[bb=31 428 583 569,width=2.0\columnwidth]{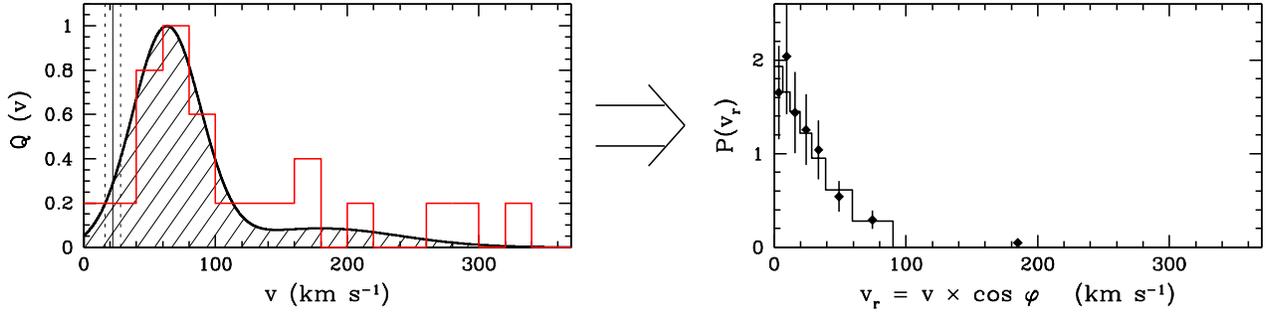}
\end{center}
\caption{
(left) Best-fit distribution of space velocities Q($v$) consisting of 
a more-likely low-velocity and a less-likely high-velocity component.
The vertical lines mark the median and 1-$\sigma$ standard
deviation of the expansion velocity of their parent PNe.  
The red histogram corresponds to the velocity distribution of 
jets in PNe whose space velocity has been determined using 
different methods and assumptions (see Table~2). 
(right) Observed distribution of radial velocities of
the jets P($v_{\rm r}$) in our sample (dots) and best-fit
synthetic distribution (black solid histogram).  
The errorbars on the observed distribution of $v_{\rm r}$ are assumed 
to be the square root of the number of jets per bin.
Bin widths are selected to have similar number of jets per bin.  
}
\label{vr_fit}
\end{figure*}

Let us suppose that all jets in PNe expanded with identical
velocity $v_0$ (top-left panel of Figure~~\ref{vr_theory}).  
The distribution of radial velocities $v_{\rm r}$ is obtained by
considering the above space distribution for each $v_{\rm r}$
interval.
The probability to detect it with radial velocity $v_1 < v_{\rm r} < v_2$
is obtained by integrating the isotropic space distribution $f(\varphi,\psi)$ 
for all the values of $\varphi$ and $\psi$ such that the projected radial 
velocity, as defined by Eq.~\ref{eq.vr}, is included in the velocity bin
from $v_1$ to $v_2$.
This can be written as:
\begin{equation}
  P(v_1:v_2) =
  \int_{\varphi_1}^{\varphi_2}\int_{\psi_1}^{\psi_2} f(\varphi,\psi) d\psi \sin\varphi d\varphi
\end{equation}
and resolved as: 
\begin{equation}
P(v_1:v_2) =
\frac{1}{4 \pi} \int_{\varphi_1}^{\varphi_2}\int_{\psi_1}^{\psi_2} d\psi \sin\varphi d\varphi.  
\end{equation}
Since we can write
\begin{equation}
  dv_{\rm r} = -v_0 \sin\varphi d\varphi \; \rightarrow \; 
  \sin\varphi d\varphi = \frac{-dv_{\rm r}}{v_0}
\end{equation}
and because $v_{\rm r}$ does not depends on the azimuthal angle $\psi$, 
the above integral results in
\begin{equation}
  P(v_1:v_2) = \frac{1}{4 \pi} \int_{\varphi_1}^{\varphi_2} 2 \pi \sin\varphi d\varphi 
  = \frac{1}{2}\int_{v_2}^{v_1} \frac{-dv_{\rm r}}{v_0}
\end{equation}
which is reduced to
\begin{equation}
  P(v_1:v_2) = \frac{1}{2} \frac{v_2-v_1}{v_0}.  
\end{equation}
Actually, since we are considering the absolute value of $v_{\rm r}$,
we should add both the P($v_1:v_2$) and P($-v_2:-v_1$) terms, then
resulting
\begin{equation}
  P(|v_1|:|v_2|) = P(v_1:v_2) + P(-v_2:-v_1) = \frac{v_2-v_1}{v_0}.
  \label{eq_probvr}
\end{equation}

Equation \ref{eq_probvr} implies that it is equally probable
to detect a jet whose space velocity is $v_0$ (top left panel
of Fig.~\ref{vr_theory}) with any value of the absolute radial
velocity $v_{\rm r} < v_0$ (top right panel of the same figure).

Certainly the Q($v$) distribution of velocities can be expected
to be different from the single velocity considered above.  
In such a case, the probability P($v_{\rm r}$) to detect a jet 
at a radial velocity $v_{\rm r}$ will be the contribution of the 
probabilities of all outflows with space velocities $v>v_{\rm r}$.
For instance, for the top-flat Q($v$) distribution between two velocities
shown in the left-middle panel of Figure~\ref{vr_theory}, the observed
distribution of $v_{\rm r}$ would be that shown in the right-middle panel
of this figure.  
A very similar result is obtained if using a Gaussian distribution for 
the space velocity rather than a top-flat distribution, as also shown 
in the middle panels of Figure~\ref{vr_theory}.

\subsubsection{Best-fit Space Velocity Distribution}
\label{subsubsec.best_vel}

The shapes of the above distributions do not match that of the 
observed  P($v_{\rm r}$) distribution shown in the top panel of
Figure~\ref{histo}.  
Different initial distributions of space velocity Q($v$) have been tested
to find that the simplest one producing a distribution of radial velocities
P($v_{\rm r}$) similar to that observed consists of two (top-flat or Gaussian)
components, one more likely at low velocities to fit the bulk of jets with
$v_{\rm r} \leq 100$ km~s$^{-1}$ and another less frequent high-velocity
component to reproduce the high-velocity tail in P($v_{\rm r}$).  
Such Q($v$) distributions and the resulting P($v_{\rm r}$) distributions 
are illustrated in the bottom-left panel of Figure~\ref{vr_theory}.

Since the detailed shape (top-flat or Gaussian) of these components in Q($v$)
results in very little differences in the resulting P($v_{\rm r}$) distributions,
a space velocity distribution Q($v$) represented by two Gaussian components
with mean spatial velocities $V_1$ and $V_2$, standard deviations $\sigma_1$ 
and $\sigma_2$, and probability ratio $\alpha={\rm P}_2/{\rm P}_1$ has been
adopted in order to determine the parameters of Q($v$) that best fit the 
observed P($v_{\rm r}$) distribution.
The jets in Table~\ref{tab:charac1} have been distributed among eight
bins with similar number of jets ($\simeq$12) but different velocity
range to allow a statistically unbiased fit.  
We note, however, that the number of outflows with velocities
$\geq$100 km~s$^{-1}$ is too small (only 10 out of 85) to provide
an adequate constrain to the high velocity component of Q($v$).  
Instead, we have assumed this high velocity component to have a mean
velocity $V_2 = 180$ km~s$^{-1}$ and a width $\sigma_2 = 60$ km~s$^{-1}$.  
The parameters of the low velocity component of Q($v$) and the probability
ratio between the two components have been sampled and a synthetic
P($v_{\rm r}$) computed for each set of $V_1$, $\sigma_1$, and $\alpha$.  
The number of jets per velocity bin in this synthetic
distribution is then compared to those in the observed
distribution P($v_{\rm r}$) and the difference minimized
using a $\chi^2$ optimization algorithm.
The best fit Q($v$), with a reduced $\chi^2 \simeq 0.9$, is shown in
Figure~\ref{vr_fit}-{\it left}, with the resulting P($v_{\rm r}$)
in Figure~\ref{vr_fit}-{\it right} shown as a solid black histogram.  
The space velocity of most jets ($\simeq$70\%) can be
described by the low-velocity Gaussian in the right
panel of Figure~\ref{vr_fit} with mean velocity 66
km~s$^{-1}$ and standard deviation 26 km~s$^{-1}$
($FWHM \simeq 60$ km~s$^{-1}$).

\begin{deluxetable}{lrl}
\tablenum{2}
\tabletypesize{\scriptsize}  
\tablecaption{Space Velocity of Selected PNe}
\tablecolumns{3}
\tablewidth{0pt}
\tablehead{
\multicolumn{1}{l}{Name}  &
\multicolumn{1}{c}{$v$}   & 
\multicolumn{1}{l}{Reference} \\ 
\colhead{} & 
\multicolumn{1}{c}{(km~s$^{-1}$)} &
\colhead{} 
}
\startdata
Abell\,63  & 126~~~~~ & \citet{Mitchell_etal2007} \\
Fg\,1      &  85~~~~~ & \citet{LMP1993} \\
Hb\,4      & 220~~~~~ & \citet{LSM97} \\
Hu\,1-2    &  72~~~~~ & \citet{M01} \\
Hu\,2-1    & 340~~~~~ & \citet{MBGR2012} \\
HuBi\,1    & 180~~~~~ & Guerrero et al., in prep.\ \\
IC\,4634   & 300~~~~~ & \citet{GMRetal2008} \\
IC\,4846   & 145~~~~~ & \citet{MGT01} \\
K\,1-2     &  35~~~~~ & \citet{AG2016} \\
Necklace   & 105~~~~~ & \citet{CSMetal2011} \\
NGC\,6572  & 46,60~~~~~ & \citet{AG2016} \\
NGC\,6751  & 79~~~~~ & \citet{CGLetal2010} \\
NGC\,6778  & 270,460~~~~~ & \citet{GM2012} \\
NGC\,6881  &  12~~~~~ & \citet{GM98} \\
NGC\,6884  &  55~~~~~ & \citet{MGT99} \\
NGC\,6891  &  80~~~~~ & \citet{GMMV00} \\
NGC\,7354  &  60~~~~~ & \citet{CVM2010} \\
M\,1-32    & 180~~~~~ & \citet{RG17} \\
M\,2-42    &  70~~~~~ & \citet{AL2012} \\
M\,2-48    &  95~~~~~ & \citet{LLEetal2002} \\
M\,3-15    & 100~~~~~ & \citet{AL2012} \\
Wray\,17-1 &  70~~~~~ & \citet{AG2016} \\
\enddata
\label{tab:true}
\end{deluxetable}

We noted in \S\ref{sec:intro} that the space velocity of 
the jets of a number of PNe had been derived on the basis of 
their shock-velocity or spatio-kinematic models under 
``reasonable'' assumptions for their inclination with the line 
of sight.  
These space velocities have been compiled in Table~2 
and their velocity distribution 
shown as a red histogram on the left panel of Figure~\ref{vr_fit}.  
There is a noticeable correspondance between the distribution 
of space velocities Q($v$) derived in this section and 
the distribution as measured in individual PNe, with a prevalence 
$\simeq$40\% for the detection of faster jets in the latter, most
likely revealing that the spatio-kinematic modeling or the 
determination of shock-velocities is favored for faster jets.

\subsection{Distribution of the Projected Radial Distances to the CSPN of Jets}

The linear distance of a jet to its CSPN ($r$) is related
with its angular distance to the CSPN ($\theta$) and the
distance to the PN ($d$) as
\begin{equation}
r = \tan \theta \times d \simeq \theta \times d.   
\end{equation}
Actually, the parameters directly available to us are their angular
and linear distances to the CSPN projected onto the plane of the sky,
$\theta_s$ and $r_{\rm s}$, respectively 
\begin{equation}
  r_{\rm s} \simeq \theta_s \times d
           = \theta \times d \times \sin \varphi 
           \simeq r \times \sin \varphi,  
\label{eq.rs}    
\end{equation}
where $\varphi$ is the inclination angle of the
jet with the line of sight as defined in \S3.1.1.  
The distribution of $r_{\rm s}$ in the middle panel of Figure~\ref{histo}
shows the number of jets per bin of distance to the CSPN
projected onto the plane of the sky.  
The distribution of $r_{\rm s}$ peaks at 0.05--0.15 pc and falls 
gradualy up to 0.4 pc, with only 12 few jets (14\%) at larger
distances than 0.4 pc from the CSPN and even fewer, only seven
(8\%), at distances smaller than 0.05 pc.

\begin{figure*}
\begin{center}
\includegraphics[bb=31 318 583 711,width=1.85\columnwidth]{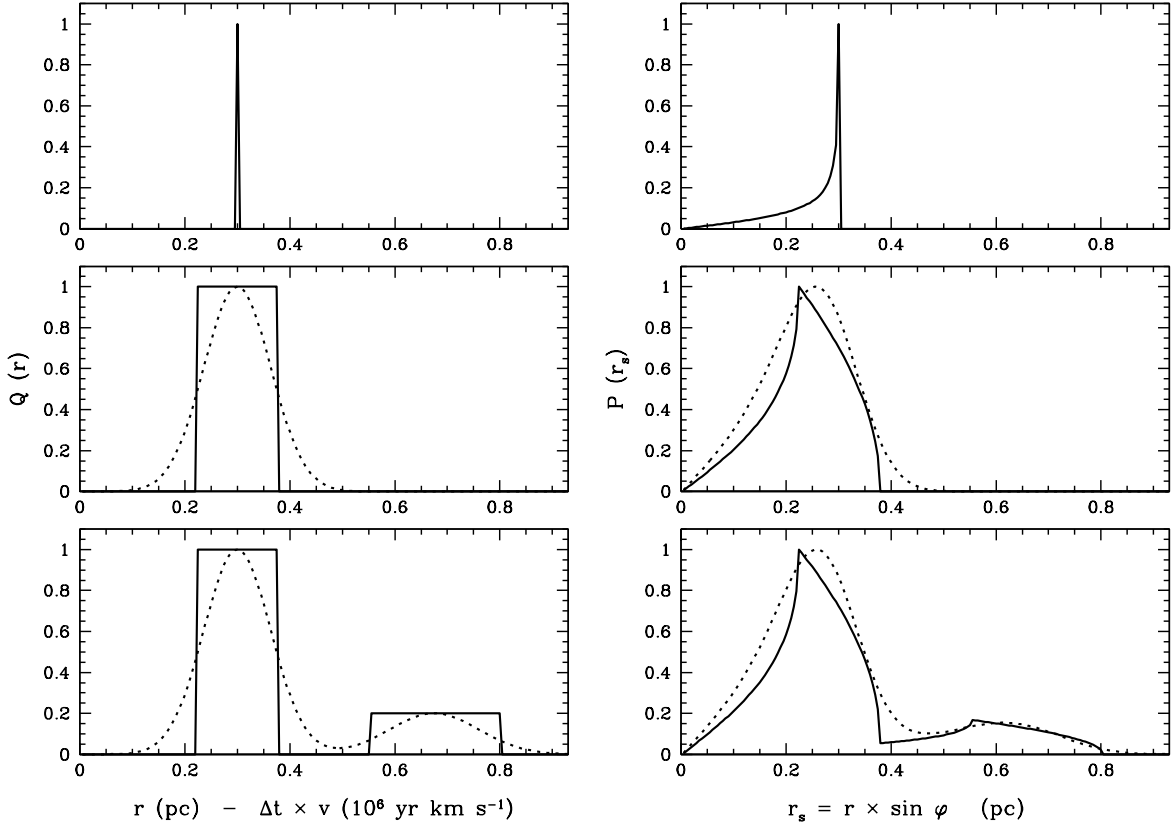}
\end{center}
\caption{
(\emph{left}) Different distributions of actual linear distances Q($r$) and 
(\emph{right}) their corresponding distribution of projected distances onto
the plane of the sky P($r_{\rm s}$).  
The Q($r$) distribution consists of a single value of 0.3 pc for $r$
in the top panel, a single top-flat (solid line) and Gaussian (dashed
line) component centered at $r$=0.3 pc in the middle panel, and two
top-flat (solid line) and Gaussian (dashed line) components centered
at $r$=0.3 pc and $r$=0.65 pc in the bottom panels.  
In the latter, the larger distance component has lower
probability than the smaller distance component.  
}
\label{x_theory}
\end{figure*}

\subsubsection{Implications for the distribution of linear distances to the CSPN}


Let us define the distribution of the linear distances (i.e., unprojected) 
of jets to their CSPNe $r$ as Q($r$) and that of the observed distribution
of distances projected onto the plane of the sky $r_{\rm s}$ as P($r_{\rm s}$).  
It is possible to derive information on Q($r$) from P($r_{\rm s}$)
if we adopt the same assumptions on the isotropic distribution of
jets and its unbiased detection.
The distribution of sky projected distances $r_{\rm s}$ is then obtained by
considering the space distribution in Equation~\ref{eq_iso} for each $r_{\rm s}$
interval.
The probability to detect it at a radial distance $r_1 < r_{\rm s} < r_2$
is obtained by integrating the isotropic space distribution $f(\varphi,\psi)$ 
for all the values of $\varphi$ and $\psi$ such that the projected distance 
$r_{\rm s}$ as defined by Eq.~\ref{eq.rs} 
is included in the radial distance bin from $r_1$ to $r_2$.
This can be written as:
\begin{equation}
  P(r_1:r_2) = \int_{\varphi_1}^{\varphi_2}\int_{\psi_1}^{\psi_2} f(\varphi,\psi) d\psi \sin\varphi d\varphi
\end{equation}
and resolved as
\begin{equation}
  P(r_1:r_2) = \frac{1}{4 \pi} \int_{\varphi_1}^{\varphi_2}\int_{\psi_1}^{\psi_2} d\psi \sin \varphi d\varphi.  
\end{equation}
We can write
\begin{equation}
  dr_{\rm s} = r \cos\varphi d\varphi \; \rightarrow \; 
  d\varphi = \frac{dr_{\rm s}}{r \, \cos\varphi}
\end{equation}
and because $r_{\rm s}$ does not depends on the azimuthal angle $\psi$, 
the above integral results in
\begin{equation}
  P(r_1:r_2) = \frac{1}{4 \pi} \int_{\varphi_1}^{\varphi_2} 2 \pi \sin\varphi d\varphi 
  = \frac{1}{2r} \int_{r_1}^{r_2} \sin\varphi \frac{dr_{\rm s}}{\cos\varphi}.
\end{equation}
Since
\begin{equation}
  \frac{\sin\varphi}{\cos\varphi} =
  \frac{r_{\rm s}/r}{\sqrt{r^2-r_{\rm s}^2}/r} =
  \frac{r_{\rm s}}{\sqrt{r^2-r_{\rm s}^2}}
\end{equation}
then the above integral can be solved as
\begin{equation}
  P(r_1:r_2) = \frac{1}{2} \frac{1}{r} \int_{r_1}^{r_2}
  \frac{r_{\rm s}}{\sqrt{r^2-r_{\rm s}^2}} dr_{\rm s} =
  \frac{\sqrt{r^2-r_1^2}}{r} - \frac{\sqrt{r^2-r_2^2}}{r}.  
\end{equation}

The behavior of this result is shown in Figure~\ref{x_theory} for different
distributions Q($r$) of the unprojected radial distances $r$ of the jets.  
If we adopted a single $r_0$ value for all of them (left-top
panel of Figure~\ref{x_theory}), then the expected P($r_{\rm s}$)
distribution (right-top panel of Figure~\ref{x_theory}) reveals
that it is more likely to detect jets at projected distances
close to their actual distances $r_{\rm s} \simeq r_0$. 
Like in Section~3.1.1 and Figure~\ref{vr_theory}, we have assumed similar
single and double top-flat and Gaussian distributions of Q($r$), as shown
in the middle and bottom left panels of Figure~\ref{x_theory}.
The resulting P($r_{\rm s}$) distributions are plotted in the corresponding
right panels of Figure~\ref{x_theory}.

\begin{figure*}[ht!]
\begin{center}
\includegraphics[bb=31 428 583 569,width=2.0\columnwidth]{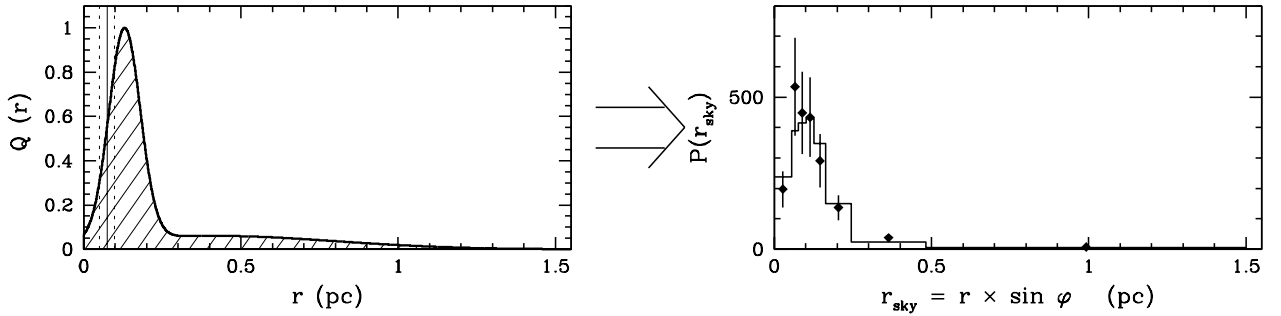}
\end{center}
\caption{
(left) Best-fit distribution of distances of jets to their CSPNe 
Q($r$) consisting of a more-likely closer and a less-likely farther 
component.  
The vertical lines mark the median and 1-$\sigma$
standard deviation of the linear size of their parent PNe.  
(right) Observed distribution of distances projected onto the plane of 
the sky of the jets P($r_{\rm s}$) in our sample (dots) 
and best-fit synthetic distribution (black solid histogram).  
The errorbars on the observed distribution of $r_{\rm s}$ are assumed 
to be the square root of the number of jets per bin.  
Bin widths are selected to have similar number of jets per bin.  
}
\label{r_fit}
\end{figure*}

\subsubsection{Best Fit Model}
\label{subsubsec.best_size}

Like the observed $v_{\rm r}$ distribution, the observed $r_{\rm s}$
distribution P($r_{\rm s}$) suggests that the distribution of actual
distances Q($r$) of jets in PNe needs to be modeled using one
component for small ($\leq$0.3 pc) values of $r_{\rm s}$ and a
second one for large values of $r_{\rm s}$.  
Since there are very little differences between top-flat and Gaussian shaped
distributions, we will assume that Q($r$) consists of two Gaussian components
with mean values $R_1$ and $R_2$, standard deviations $\sigma_1$ and $\sigma_2$,
and probability ratio $\alpha={\rm P}_2/{\rm P}_1$.

\begin{figure*}
\begin{center}
\includegraphics[bb=60 190 565 530,width=1.85\columnwidth]{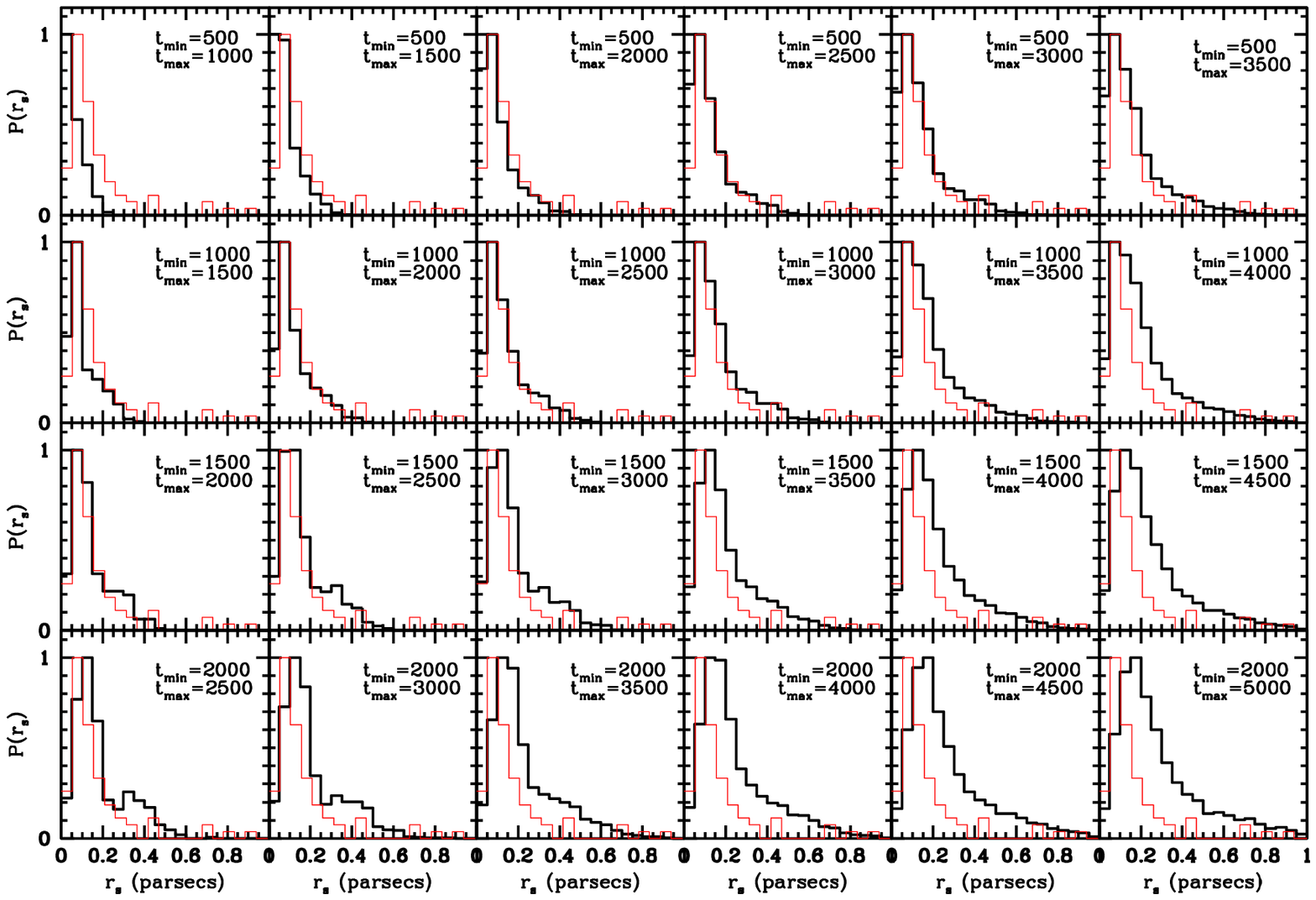}
\end{center}
\caption{
Monte Carlo simulated distributions of P($r_{\rm s}$) for a range of values of 
the time span ($t_{\rm max}$) and minimum age ($t_{\rm min}$) of a population 
of collimated outflows (black histogram) compared to the observed P($r_{\rm s}$) 
distribution (red histogram).  
The best matches between the synthetic and observed P($r_{\rm s}$) 
distributions are found for 1,500~yrs~$\leq t_{\rm max}\leq$~3,000 yrs, 
and 500~yrs~$\leq t_{\rm min}\leq$~1,500 yrs.  
}
\label{fig.monte_carlo}
\end{figure*}

The jets in Table~\ref{tab:charac1} have been distributed among eight
bins with similar number of outflows but different widths to allow
a statistically unbiased fit.  
The best-fit parameters have also been obtained using a $\chi^2$
optimization algorithm comparing the synthetic P($r_{\rm s}$) for a space
of reasonable parameters and the observed distribution of distances to
the CSPN projected onto the plane of the sky shown in the middle panel of
Figure~\ref{histo}.  
The best-fit has a reduced $\chi^2 \simeq 0.9$ and the results 
are shown in Figure~\ref{r_fit}.  
The best-fit for P($r_{\rm s}$) is achieved with a Q($r$) distribution
consisting of a Gaussian component with mean radius 0.135 pc and
standard deviation 0.055 pc, and another one with mean radius 0.4 pc
and standard deviation 0.4 pc.
Most jets ($\simeq$70\%) can be adscribed to the first component.

\subsection{Age, Time Span, and Formation Time of Jets} 
\label{subsec.times}

The velocity and spatial distributions of jets derived in 
\S\ref{subsubsec.best_vel} and \S\ref{subsubsec.best_size} 
are not independent.
These correlations depend on the distributions of the age and 
time span of the jets (i.e., their duration before they disperse), 
and on dynamical effects modifying their kinematics.
Therefore, it is possible to estimate the time span ($t_{\rm max}$) and minimum 
age ($t_{\rm min}$) of the jets in our sample by comparing the observed spatial 
distributions of outflows P($r_{\rm s}$) and that derived from their
distribution of space velocities Q($v$).  
The latter has been built using Monte Carlo simulations of a sample of
jets whose space velocities are described by the Q($v$) distribution
derived in \S\ref{subsubsec.best_vel}.  
In these simulations, jets have been shot along random directions
every year for a total number of $5\times10^5$ outflows to assure
that the system achieves a stationary regime, i.e., even the lowest
speed outflows could fully evolve until they dissolve after a time
span $t_{\rm max}$.

Different synthetic P($r_{\rm s}$) distributions resulting from simulations
exploring a range of values of $t_{\rm max}$ and $t_{\rm min}$ are presented
in Figure~\ref{fig.monte_carlo}, together with the observed P($r_{\rm s}$)
distribution.  
An inspection of this figure allows us to constrain the
values of $t_{\rm max}$ and $t_{\rm min}$.  
The minimum age $t_{\rm min}$ corresponds to the time lapse between the
ejection of the jet and the present time and thus, as it increases, the
distribution shifts towards larger values of $r_{\rm s}$.
Synthetic P($r_{\rm s}$) distributions for minimum ages 
$t_{\rm min} \lesssim 500$ yrs or $t_{\rm min} \gtrsim 1,500$
yrs cannot reproduce the observed distribution of $r_{\rm s}$.
On the other hand, the time span $t_{\rm max}$ determines the maximum
distance traveled by a jet before it disperses and thus, as it
increases, the synthetic distribution broadens and reaches further
away from the CSPN.  
Synthetic P($r_{\rm s}$) distributions for time spans in excess of 3,000 yrs
(shorter than 2,000 yrs) are broader (narrower) or peak at larger (smaller)
linear distances to the CSPN than the observed distribution.  
The minimum age of 1,000$\pm$500 yrs and time span of 2,500$\pm$500 yrs 
derived above mark the lower and upper limits, respectively, for the 
distribution of ages of jets in this sample of PNe.

Unlike studies of jets in individual PNe, where the inclination angles are
unknown, the distribution of kinematic ages of the jets in our sample can
be derived from their distributions of space velocities Q($v$) and actual
distances to the CSPNe Q($r$).  
A fraction $\simeq$70\% of the outflows have ``low-speed'' with a mean
space velocity of 66 km~s$^{-1}$ and another fraction $\simeq$70\% are
``small'' with mean distance to the CSPN of 0.135 pc.  
Therefore, at least 50\% of the outflows in our sample (i.e., 
0.70$\times$0.70) have ``low-speed'' and are ``small'', and 
their average age is $\sim$2,000 yrs, with a 1-$\sigma$ width 
of $\pm$1,700 yrs according to the 1-$\sigma$ widths of Q($v$) 
and Q($r$).  
Interestingly, a very similar age $\sim$2,200 yrs is derived for ``high-speed''
and ``large'' outflows adopting an average distance to the CSPN of 0.4 pc and a
mean expansion velocity of 180 km~s$^{-1}$.  
These mean values for the ages of the jets in this sample of PNe 
are fully consistent with their minimum ages and time spans.

Furthermore, the average age of the jets in this sample and their minimum 
age and time span are consistent with the age of their parent PNe, which 
was estimated in \S\ref{subsec:sample1} to be 1,000--3,000 yrs.  
We need to keep in mind that this sample, as described in
Section~\ref{sec:sample}, does not include proto-PNe, and 
then very young outflows are not considered.  
Including additional jets located near the CSPN in our sample would presumably
add also younger PNe, thus reducing both the minimum age $t_{\rm min}$ of the
jets and the mean age of their parent PNe.  
At any rate, this suggests that a noticeable fraction of jets 
and their parent PNe are mostly coeval, i.e., they formed at 
about the same time.  
This is a similar result to that reported by \citet{Huggins2007}, who
investigated the age of jets and tori in a small sample of proto-PNe
and found that their occurrence is almost simultaneous, but it conflicts 
results implying that jets are launched a few thousand years before or
after the common-envelope interaction that formed the PN itself
\citep{TMW2014}.  
Those results, however, are based on a small sample of only four 
post-common-envelope PNe that, in spite of the addition of a few 
more recent determinations \citep{JB2017,Derlopa_etal2019}, are 
affected by the adoption of kinematic ages as nebular ages 
(see the discussion on this issue in \S\ref{subsec:sample1}).

The coeval formation of the jets and their PNe helps us interpret the 
observed distribution of the relative distance $x$ with respect to the 
nebular radius along the direction of the jets (bottom panel of
Fig.~\ref{histo}).
Since jets and PNe were formed at the same time and jets have typical 
expansion velocities $\simeq$3 times faster than those of their parent 
PNe, it is unlikely to detect a jet close to the CSPN.  
Indeed, it is conspicuous the small number of jets 
detected in the close proximity of CSPNe;  only
the two outflows detected spectroscopically in
HuBi\,1 and M\,1-32 are found below half the nebular
radius $x<0.5$.  
On the contrary, jets can be expected to be found mostly at 
a value of $x$ close to three nebular radii, but most have 
relative distances in the range $0.5 \leq x \leq 3$, with a 
notable peak in the bins in the range $0.75 \leq x \leq 1.5$, 
and only a small fraction, $\eqsim20\%$, are at relative 
distances at least three times larger than the nebular radius. 
The large fraction of jets with values of $x \simeq 1$ suggests that 
the jet was not only ejected by the same time the PN formed, but that 
jet and PN expanded together as an ensemble.

\section{Two Populations of Jets}
\label{sec.disc}

The analyses of the distributions of the observed radial velocity and
projected distances to the CSPN of the PN jets have both yielded
bimodal distributions for the space velocity and unprojected distance
to the CSPN. 
The radial distribution of the jets can be inferred from their
velocity distribution for reasonable values of their time span
and age.  
Accordingly, jets in PNe can be adscribed to two different populations:
most (70\%) are ``low-speed'' (mostly at ``small'' distances from the CSPN)
with mean space velocities $\sim$66$\pm$26 km~s$^{-1}$, 
and
a smaller fraction (30\%) are ``high-speed'' (mostly at ``large'' distances
from the CSPN) with space velocities $>$100 km~s$^{-1}$.

\citet{GCM01} concluded that the low-ionization structures of PNe, of
which jets are a prominent group, exhibit a wide range of properties.  
The bimodal distribution of space velocities of jets in PNe
derived in \S\ref{subsubsec.best_vel} reinforces their
suggestion on a different nature among low- and high-velocity
outflows.
Their distinct kinematic properties most likely imply different
launching mechanisms resulting in different prescriptions for their
space velocities and/or the way they interact with the nebular 
envelope.  
There is a wealth of literature discussing the formation and launch 
of jets in PNe, but they can be basically split into two major scenarios.

In a first scenario, jets may result from hydro-dynamical 
effects that focus initially isotropic fast winds through
oblique shocks associated with a non-spherical low-velocity
AGB wind \citep{Balick87,BPI87,FBL96}.
In all these models, outflows arise at noticeable distances from the CSPN.  
Typical examples of this collimation process are the FLIERs
detected in NGC\,6826 and NGC\,7009, where the outflows are
found at the tips of an elliptical inner shell and connected
to these by some filamentary emission \citep{BPI87}.  
Very recently, \citet{BFL2019b} have explored the additional effects of a
toroidal magnetic field on tapered (non isotropic) fast stellar winds and
concluded that jets and bullet-like features may form with a range of
expansion velocities, depending on the initial injection speed of the fast
stellar wind in their models.
Somehow related is the collimation of outflows through the interaction 
of a fast wind with a wrapped disk \citep{Retal05} that result in low 
velocity jets, with typical velocities of a few tenths of km~s$^{-1}$.

More usually, jets in PNe are accepted to be the result of ballistic 
ejections, when two compact blobs (``bullets'') are symmetrically
ejected at high speed from the CSPN \citep{Soker90}.
Magnetic fields have been proposed to be at the origin of
this ``bipolar'' ejection \citep{GS1999,Steffen2009}, but
it is unlikely their effects become noticeable in single
stars \citep{GS2014}.  
Accordingly, the focus has rather turned into the interaction of the precursor
AGB star with a stellar or sub-stellar companion
\citep{Livio82,SL94,Soker96,HarpazSoker94}, with the possible contribution of
magnetic fields emanating from the accretion disk
\citep{FB2004,Dennis_etal2009}.
Observational evidence for the existence of circumbinary
disks and accretion disks around dwarf companions
\citep{Bujarrabal_etal2013,Bollen_etal2017,Ertel_etal2019}
has lent strong support to the formation of jets in binary
systems.
The outflow would form close to the CSPN, in line with the
results of \citet{Huggins2007}, who found collimated outflows
as close as 9$\times$10$^{15}$ cm (0.003 pc) in a small sample
of proto-PNe.
The expansion velocity and mass of these bullets will depend on the
orbital parameters, accretion rate, and properties of the secondary
star.  
It must be noted that the escape velocity for a 1 $M_\odot$ main-sequence
of dwarf companion is much higher \citep[e.g.][]{Bollen_etal2019} than the
typical jet space velocity derived in this work.

The theoretical models for jet formation presented above have their
difficulties to predict the low-velocity jets mostly detected in PNe,
unless \emph{``ad hoc''} low velocities are assumed for the stellar
wind or the responsible for the jet collimation is an accretion disk
around a sub-stellar companion.
Alternatively, the interaction of a jet with the previous slow
AGB wind can certainly modify its initial velocity, slowing it
down.  
Indeed, among the ``low-speed'' component in the Q($v$) distribution derived
in \S\ref{subsubsec.best_vel}, there is a non-negligible fraction of jets
with expansion velocities similar or smaller than the median expansion 
velocities of their parent PNe (Figure~\ref{vr_fit}-\emph{left}).  
These can be identified with the low velocity low-ionization 
features described by \citet{Perinotto_etal2004} as SLOWERs 
or by \citet{GCM01} as jetlike, although it must be noted that
those definitions are based on observed radial velocities and
thus, unlike the leftmost tail of the Q($v$) distribution shown
in the left panel of Figure~\ref{vr_fit}, they are subject to
notable corrections by their unknown inclination with the line
of sight.  
The interaction between a jet and the nebular envelope is explored below.
Two broad cases can be envisaged depending on whether the
outflows can be considered heavy or light \citep{BFL2019}.

``Heavy'' outflows are those with a linear momentum density
much larger than the AGB wind, and thus they can be expected
to experience very little braking in their interaction with
the nebular envelope.
Heavy jets may then result in a ``high-speed'' 
population of jets with expansion velocity in
excess of a few hundreds km~s$^{-1}$. 
Only a handful PNe present jets with radial velocities $>$100 km~s$^{-1}$,
namely Hb\,4, Hen\,2-186, HuBi\,1, KjPn\,8, M\,1-16, M\,1-32, MyCn\,18,
NGC\,2392, NGC\,6778, although there might be certainly additional
``high-speed'' jets that are not directly identified because projection
effects decrease their radial velocities.
These are pressumably the progeny of young-PNe and proto-PNe with the fastest
jets, such as CRL\,618 \citep{Lee13} and Hen\,3-1475 \citep{BH01,RGM03}, or
water-fountains such as IRAS\,18113$-$2503 \citep{Orosz_etal2019}. 
Interestingly, some of the PNe with the fastest
jets are known to harbor binary systems 
\citep[e.g., MyCn\,18 and NGC\,2392,][]{Miszalski_etal2018,Miszalski_etal2019} 
or have gone through a common-envelope phase 
\citep[e.g., NGC\,6778,][]{Miszalski_etal2011}.

``Light'' outflows are those whose density is much lower than that of the
slow AGB wind, according to the definition introduced by \citet{AS2008}.
Dynamical effects will  certainly distort the space and velocity
distributions of ``light'' jets produced at the end of the AGB
and onset of the proto-PN phases as they interact with the slow
AGB wind \citep{ST98}.  
This interaction will slow down their expansion velocities, 
resulting in a ``low-speed'' population of jets.
In the most extreme cases of light and wide opening angle outflows, they can
get trapped within the nebular envelopes as regions of enhanced density that
expand with the PN \citep{BFL2019}.  
The subsequent interaction of this imprint on the AGB wind with the fast
post-AGB wind and strong D-type ionization fronts would shape the
PN and may even speed up these features as they are entrained with the
nebular envelope \citep{ST98,Soker2002,LS03,Huarte2012}.
These would correspond with the significant fraction of outflows
that expand at similar expansion velocities than their parent PNe
and whose projected positions onto the main nebular shell suggest
they are embedded within it.

\section{Conclusions}

We have presented an analysis of the kinematic and spatial properties
of a sample of 85 jets detected in 58 PNe.
By considering the probability to detect outflows with different
radial velocities and at different radial distances projected onto
the plane of the sky, the distributions of their inclination angle
independent space velocity and actual distance to the CSPN have
been derived.
These happen to be bimodal in both cases.
The main component (70\%) of the distribution of the space velocity can be
described by a Gaussian function with mean velocity of 66 km~s$^{-1}$ and
1-$\sigma$ dispersion of 26 km~s$^{-1}$ (or $FWHM \simeq 60$ km~s$^{-1}$),
whereas the secondary (30\%) component has mean velocity of 180 km~s$^{-1}$ and
1-$\sigma$ dispersion of 60 km~s$^{-1}$ (or $FWHM \simeq 140$ km~s$^{-1}$).  
Similarly, the main component (70\%) of the distribution of the
actual (unprojected) distance to the CSPN can be described by a
Gaussian function with mean distance 0.135 pc and 1-$\sigma$
dispersion of 0.055 pc (or $FWHM \simeq 0.13$ pc), whereas the
secondary (30\%) component has mean distance of 0.4 pc and
1-$\sigma$ dispersion of 0.4 pc (or $FWHM \simeq 1.0$ pc).

The comparison of the observed spatial distribution of jets
and that derived from their velocity distribution implies
that jets are mostly coeval with the formation of their parent
PNe and that they disperse in time-scales shorter than
$\simeq$2,500 yrs.  
Since jets are coeval to their parent PNe and they last for no more than
2,500 yrs, PNe that harbor jets are young, with ages $\lesssim$3,000 yrs.

The low space velocity of jets among a significant fraction of PNe poses
notable difficulties for models of jet formation, particularly for those
where jets are formed in an accretion disk around a stellar companion.  
This needs to be linked with the significant number of jets that
are found close to the edge of the nebular envelopes, traveling
away at velocities similar to the expansion velocities of their
parent PNe.  
These facts support the prevalence among detected jets of ``light'' outflows
that get trapped when interacting with the nebular envelopes, depositing their
momentum and energy to actively shape their parent PNe.
A possible bias against fast jets at large
distances from their PNe cannot be ignored.

\acknowledgments

The authors acknowledge support of the grants AYA 2014-57280-P and
PGC2018-102184-B-I00, co-funded with FEDER funds.
J.S.R.-G. acknowledges the financial support from DGAPA-PAPIIT grants
IN103117 (PI Dr.\ Miriam Pe\~na) and IG100218 (PI Dr.\ Pablo F.
Vel\'azquez) and scholarship from BECA MIXTA CONACyT-M\'exico.  
This research was motivated long time ago during
an inspiring coffee talk with Dr.\ Hugo Schwarz.  
The authors deeply appreciate Dr.\ Pablo F.\ Vel\'azquez for his 
helpful comments, Dr.\ Jos\'e A.\ L\'opez for a throughful reading
of the manuscript and providing us access to spatio-kinematic data
from the SPM Kinematic Catalogue of Galactic Planetary Nebulae,
and Drs.\ Noam Soker and B.\ Balick for their many comments and
contributions to the final version of this manuscript.

Based on observations made with the NASA/ESA Hubble Space Telescope, and
obtained from the Hubble Legacy Archive, which is a collaboration between
the Space Telescope Science Institute (STScI/NASA), the Space Telescope
European Coordinating Facility (ST-ECF/ESA) and the Canadian Astronomy
Data Centre (CADC/NRC/CSA).
This research has made also use of the HASH PN database at hashpn.space,
the SPM Kinematic Catalogue of Galactic Planetary Nebulae, the NASA’s
Astrophysics Data System, and the SIMBAD database, operated at CDS,
Strasbourg, France.

%

\vspace{5mm}
\facilities{HST(WFPC2,WFC3), NOT(ALFOSC), SPM(2.1m,MES)}

\end{document}